# The Growth of New Extended Carbon Nanophases from Ferrocene Inside Single-Walled Carbon Nanotubes

**Hans Kuzmany**[*,1], **Lei Shi**[*,1], **Jenő Kürti**[2], **János Koltai**[2], **Andrey Chuvilin**[3], **Takeshi Saito**[4], **Thomas Pichler**[1]

[1] Fakultät für Physik, Universität Wien, Strudlhofgasse 4, 1090 Wien, Austria
[2] Department of Biological Physics, Eötvös University, Pázmány P. sétány 1/A, 1117 Budapest, Hungary
[3] CIC nanoGUNE, Tolosa Hiribidea 76, 20018 Donostia-San Sebastian, Spain
  and IKERBASQUE, Basque Foundation for Science, Mara Daz de Haro 3, 48013 Bilbao, Spain
[4] Nanomaterials Research Institute, National Institute of Advanced Industrial Science and Technology (AIST), Tsukuba, Ibaraki 305-8565, Japan

* Corresponding author: e-mail hans.kuzmany@univie.ac.at, Phone: +43-1-427772607, Fax: +xx-xx-xxx
  e-mail lei.shi@univie.ac.at, Phone: +43-1-427772610, Fax: +xx-xx-xxx

The Raman response of new structures grown after filling SWCNTs with ferrocene and transformation at moderate high temperatures is demonstrated to be very strong, even stronger than the response from the tubes. Transmission electron microscopy demonstrates that the new objects are flat and exhibit a structure similar to short fragments of nanoribbons. The growth process is controlled by two different activation energies for low and high transformation temperatures, respectively. Immediately after filling Raman pattern from a precursor molecule are detected. Two different types of nanoribbons were identified by selecting special laser energies for the Raman excitation. These ribbons have the signature of quaterrylene and terrylene, respectively.

**1 Introduction** The inside of carbon nanotubes (C-NTs) provides an interesting nanospace for the growth of new materials [1]. While in the early days of this research field fullerenes were the dominating filler material, recently ferrocene (FeCp$_2$) became an interesting filler molecule due to its magnetic properties. The case of rather narrow tubes with diameters around 1 nm are special, since there is dramatic confinement for the growth process. Highly one-dimensional structures are expected to grow with enhanced interaction of the grown species and the tube walls. For these small diameter tubes ferrocene and related carbon rich molecules with typical diameters of only 0.5 nm have been used often if carbonaceous molecules are intended to be grown. We and others had recently extensively discussed the observation of new structures inside the small diameter CNTs if the filled tubes were subjected to a moderate high temperature annealing process [2–4]. Such experiments were performed mainly by Raman scattering (RS) and resonance RS but also by mass spectroscopy and transmission electron microscopy. Isotope substitution has been used as an additional technique to identify the new objects grown inside the tubes [5]. In addition quantum-chemical calculations were used, mostly on the DFT level with B3LYP hybrid functionals, to characterize the grown objects [6].

Besides the growth of inner tubes during the annealing process new non-dispersive Raman lines were observed in the frequency region of the D line now generally assigned as $C_n$ lines. These lines are located at 1245 ($C_1$), 1275 ($C_2$), and 1362 ($C_3$) cm$^{-1}$. An additional rather strong

Raman line has been observed around 470 cm$^{-1}$ and assigned as C$_{470}$. These lines exhibited strong resonance enhancement for red laser excitation with a peak resonance at 1.99 eV. This resonance was assigned as outgoing for an electronic gap energy of 1.85 eV. From experiments with deuterated ferrocene, the lines were shown to be not intrinsic to the tube system [7] but rather originate from special hydrocarbon molecules grown during the transformation process. From a comparison of these experimental data with the DFT calculations the extended new structures were suggested to be nanoribbon-like hydrocarbon molecules similar to quaterrylene or PTCDA [6]. The C$_n$ lines are in the fingerprint region of organic molecules and represent CC single bond stretching combined with CH in plane bending. The Raman line at 470 cm$^{-1}$ is a signature for ribbon-like flat objects and has been named frequently a radial breathing like mode (RBLM) [8].

Here we concentrate on the time and temperature dependent growth of the encapsulated molecules. Threshold temperatures for the start of the transformation process were found to be as low as 450 °C. High resolution transmission electron microscopy experiments showed indeed the existence of flat object inside the tubes. For the transformation process two different activation energies were observed. In the very early stages of the growth process slightly modulated Raman spectra indicate a structure which is different from the final product. Changing the growth conditions and the laser energy for excitation revealed another set of characteristic Raman lines with different fingerprint pattern.

**2 Experimental** SWCNTs were grown by the enhanced direct injection pyrolysis synthesis process (eDIPS) with mean diameters between 1.0 and 1.7 nm [9]. All tubes were purified by air treatment followed by HCl etching as reported previously [7] to remove amorphous carbon covering the catalysts and to remove catalytic particles. SWCNT buckypaper was obtained after filtrating and washing by distilled water and ethanol. In this way clean material was obtained with low D-line intensity. The tubes were opened by etching in air at 420 °C and subsequently filled with ferrocene by heat treatment together with ferrocene at 400 °C for two days in high vacuum.

Raman spectra were excited at room temperature with various lasers and recorded with a Horiba LabRAM and a Bruker RFS100/S FT-Raman spectrometer. If not otherwise specified all spectra were normalized to the intensity (height) of the 2D line.

Transmission electron microscopy (TEM) was performed on Titan 60-300 microscope (FEI) equipped with gun monochromator, imaging side Cs-corrector and Quantum GIF energy loss filter (Gatan). Images were obtained at 80 kV accelerating voltage with highly monochromated beam. For clearer presentation of internal content of the tubes the lattice of the tube walls was filtered out by application of a sharp Fourier mask.

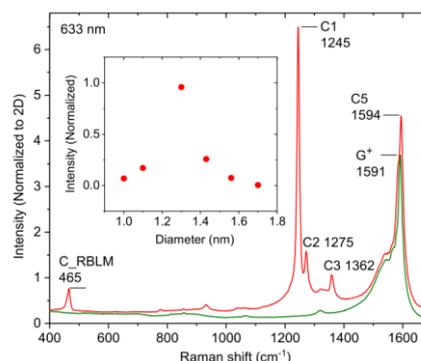

**Figure 1** Raman spectra recorded for 633 nm laser excitation of a pristine but opened DIPS tube (green) and after filling with FeCp$_2$ and transformation at 700 °C (red). C$_{RBLM}$, C$_1$, C$_2$, C$_3$, and C$_5$ assign the new lines. The D line of the tubes is between C$_2$ and C$_3$. C$_5$ and G$^+$ are almost coincident. The insert depicts relative intensities of the C$_1$ line of the new molecules grown for 1 hour transformation at 700 °C in tubes with various diameters.

For more experimental details see supplemental contribution.

**3 Results** The results presented in the following concentrate on finding optimum conditions for the growth of the transformed objects, their improved characterization, and the observation of new structures during growth.

**3.1 Optimum conditions for the growth of the species inside the tubes.** For the growth of the new compounds inside the tubes particular attention was paid to maximize their intensity in the Raman spectrum. In this way Raman intensity of the strongest line of the new molecule exceeded the intensity of the G$^+$ line of the tubes by almost a factor of two (exactly a factor of 1.87). Figure 1 depicts spectra of eDIPS tubes with 1.3 nm diameter after tube opening (green) and after filling with FeCp$_2$ and transformation for 1 hour at 700 °C (red). As will be shown below this transformation temperature is optimum for the growth of the new material. In the figure the most important Raman lines of the new molecule are assigned in the usual way as C$_1$, C$_2$, C$_3$, and C$_5$ [5] with the corresponding frequencies indicated. The radial breathing like mode is now observed at 465 cm$^{-1}$. The insert of the figure depicts observed Raman intensities of the C$_1$ line recorded after transformation of one hour at 700 °C. It demonstrates that tubes with 1.3 nm yield best results for the growth process. The values for the diameters were obtained from optical absorption. The new Raman lines can be observed for tubes as narrow as 1 nm and for tubes as wide as 1.7 nm, but for the latter cases Raman intensities are very weak. Particularly in the case of the large diameter tubes competition exists between the growth of the new molecules and the growth of inner tubes.

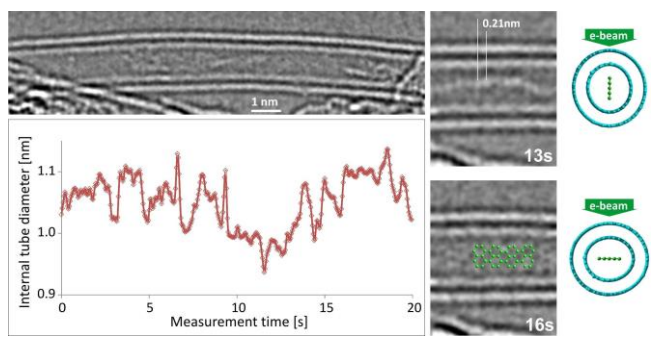

**Figure 2** Transmission electron microscopy (TEM) of double-walled carbon nanotube with molecules inside. Left top: Nanotube with included molecules. Left bottom: Fluctuation of the inner tube diameter with time. Right: snap shots of the images at two different points of time corresponding to two different molecule orientations relative to the beam. The bottom right part has an oligorylene pattern overlaid on the image of the ribbon.

### 3.2 Evidence for flat objects from TEM analysis

Evidence for flat objects inside the tubes was obtained from a smart application of TEM for the visualization of sulfur terminated classical nanoribbons [10]. In this case the molecules inside the tubes are imaged in various orientations, either with their lateral extension parallel or perpendicular to the e-beam direction. These orientations modulate the tube diameter which can be measured with high precision. A result for our case is depicted in Fig. 2. It shows the image of a carbon nanotube with an encapsulated object recorded for extended measuring time. The random modulation of the inner-tube diameter with time is the signature of a rotating flat object which causes stochastic expansion and reduction of the tube diameter. This is demonstrated in the right part of the figure which depicts snapshots of the measured inner tube diameters. For the beam orientation parallel to the molecule plane (head on geometry) only a line is observed while for the perpendicular orientation hexagonal pattern appear inside the tubes. Possible representations of the pattern are oligorylenes. The images were recorded for a tube with a rather large diameter and a double-walled structure. Wide tubes turned out to be easier to disentangle and are therefore easier to image. A very similar behavior was observed for the diameter of the outer tube. An orientation dependent modulation of the tube diameter is a clear signature for flat objects.

### 3.3 Thermal activation of the various growth processes

In order to find optimum conditions for the growth of the nanoribbons transformation of the inserted ferrocene was performed in a wide temperature range. In each case transformation was performed by exposure in high vacuum at the selected temperature for one hour. Figure 3 depicts results for the growth of the ribbons (upper) and for the growth of inner tubes (lower). For the $C_1$ line and for the $C_{RBLM}$ the normalization factor were 5.96 and 0.472, respectively. These factors give the line intensities relative to the 2D line of the nanotube. For the two selected Raman lines maximum Raman signal is obtained for transformation at 700 °C. In detail we are faced with four regions for the temperature response. In region I, up to about 450 °C almost no signal is observed indicating that there is hardly any growth of the ribbons. In the second temperature range between 450 °C and 500 °C highly activated increase of the ribbon growth is observed. Beyond 500 °C in region III the activation slows down and the signal increases slowly until it reaches a peak value for 700 °C. Interpreting the temperature dependence as an Arrhenius-type behavior allows to evaluate activation energies $Q_{II,III}$ from

$$Q = k_B * \frac{\ln(I_1/I_2)}{1/T_2 - 1/T_1}, \qquad T_2 < T_1 \qquad (1)$$

where $T_{1,2}$ are the beginning and the end of the straight lines in the regions, $I_1$ and $I_2$ the corresponding intensities, and $k_B$ is the Boltzmann constant. For details see supplemental contributions. For region II and III the activation energies are 3.9 and 0.21 eV/K, respectively. Finally, in region IV ribbon growth becomes continuously less efficient until at around 1200 °C almost no signal from the ribbons is observed.

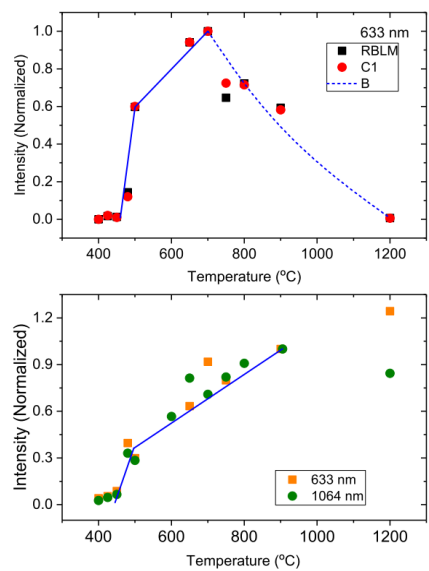

**Figure 3** Temperature dependence for the growth of nanoribbons as characterized by the area intensity of the $C_1$ (red bullets) and by the area intensity of the radial breathing like mode Raman lines (black squares) (upper). Intensities were recorded after 1 hour at the indicated temperatures. For both lines they were normalized to 1 for their highest value. Straight lines and dashed curve B are guides for the eye. The lower part of the figure depicts the intensities of the inner tubes grown during the same temperature treatment for the nominally 1.3 nm diameter tubes. Orange squares and green bullets are for excitation with the lasers as indicated and normalization to one at 900 °C.

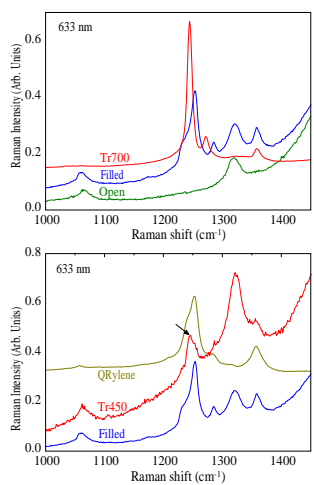

**Figure 4** Upper: Raman response for SWCNT in the fingerprint region immediately after opening (Open), after filling with ferrocene at 400 °C (Filled), and as compared to the response after optimum transformation at 700 °C (Tr700). The latter is the same spectrum as in Fig. 1 but reduced by a factor 12. The broad line at 1320 cm$^{-1}$ is the D line of the tubes. Lower: Spectra after filling (Filled), in comparison to the response from quaterrylene (QRylene), and after mild transformation at 450 °C. The dashed line is a guide for the eye. The arrow points to the position of the $C_1$ Raman line. In both figures the increasing background in the spectra originates from the low frequency tail of the G-line ensemble.

In the case of inner tube growth the data shown for the excitation with 633 nm were recorded for the intensity of the inner tube RBM at 338 cm$^{-1}$. In this case the laser was in resonance with the $E_{22}$ transition. For excitation with 1064 nm the $E_{11}$ transition was addressed. The full range of inner tube RBMs from 295-350 cm$^{-1}$ was first fitted with a set of Lorentzian lines. From the fitted data the area intensity was evaluated. As far as region I, II and III are concerned the temperature dependence of the growth process is similar as for the growth of the ribbons. The two evaluated activation energies are 2.15 and 0.18 eV/K for region II and region III, respectively. In contrast to the behavior of the ribbons inner tubes continue to grow at least up to 900 °C and do not exhibit the dramatic decrease even beyond this temperature.

An interesting behavior of the Raman response was observed right after the filling at 400 °C and after beginning of the transformation. While it was not possible to detect the response from the ferrocene molecule inside the nanotubes a spectrum very similar to but not identical to the response of the ribbons was recorded. This is demonstrated in Fig. 4 which depicts the Raman response of the tube material in the fingerprint region after opening (Open), after filling at 400 °C (Filled), and after extended transformation at 700 °C (Tr700). The spectrum (Filled) is not ferrocene which has its main Raman peak at 1107 cm$^{-1}$. It is also different from the spectrum recorded after transformation at 700 °C. The line positions for the two spectra "Filled" and "Tr700" are listed in Tab. 1, Col Filled and Col $C_n$. The table provides an overview of observed Raman lines of the new objects and related materials. Most significant differences are for the $C_1$ line at 1245 and for the $C_2$ line at 1275 cm$^{-1}$ which shift to 1253 and 1285 cm$^{-1}$, respectively.

**Table 1** Experimental and calculated frequencies and optical transition energies for the nanoribbons observed in the Raman spectra of transformed ferrocene@SWCNT and reference materials. Columns: Filled: after filling at 400 °C, QR: quaterrylene powder, $C_n$: lines for nanoribbon, $B_n$: lines for new molecule, TR: terrylene powder, QRc: calculation for quaterrylene, TRc: calculation for terrylene. Experimental transition energies are from peak resonance. Calculated values are from DFT calculations with B3LYP functionals. All frequencies are given in wave numbers. "mask" means lines are covered by the response from the tubes.

| Filled | QR | $C_n$ | $B_n$ | TR | QRc | TRc |
|---|---|---|---|---|---|---|
| mask | 1590 | 1594vs | 1607vs | 1617s | 1534vs | 1605w |
| mask | 1561 | mask | mask | 1557vs | | 1543vs |
| mask | 1546 | mask | mask | | | 1359s |
| 1358m | 1358m | 1362m | 1345vs | 1358s | 1349w | 1338vw |
| 1285s | 1285m | 1275m | 1315w | 1311w | 1274m | 1292s |
| 1253s | 1254s | 1245vs | 1260s | 1276s | 1244vs | 1257vs |
| 1238w | 1232w | | 1219vw | 1212vw | | 1202vw |
| | | 932w | | | 1056w | 1019vw |
| 552m | 550m | 465m | 413m | 537s | 540m | 525 |
| | 512w | | | | 499m | |
| transition energies | | | | | | |
| | | 1.85 eV | 2.26 eV | | 1.85 | 2.28 eV |

The lower part of Fig. 4 shows that the new spectra are almost identical to the response of quaterrylene crystalline powder (QRylene in the figure). The lines for the latter are compiled in Tab. 1, Col QR. This agreement even holds for the RBLM at low frequencies, except for the fact hat in the crystalline material all lines are Davidov split. This part of Fig. 4 shows also the possibility to follow the transition of the spectra after filling to the final spectrum of the nanoribbon by Raman spectroscopy. This is demonstrated by the spectrum labeled Tr450 which was recorded after short time transformation at 450 °C. This spectrum still shows the response from the object obtained after filling (following the dashed line) but also the upcoming $C_1$ line at 1245 cm$^{-1}$.

**3.4 A new type of nanoribbons** A set of new Raman lines was observed by changing the conditions for transformation and spectra recording. These lines result from another new object (NO) not seen so far and are named in the following as $B_n$ lines. Again, from tuning for maximum signal height the $B_n$ lines were strongest for excitation with a blue shifted laser. Strongest response was obtained for 568 nm excitation which is very close to the peak resonance at 2.26 eV. Optimum transformation temperature was 600 °C. Figure 5 depicts the spectrum after transformation at 600 °C (Tr600) in comparison to the response from the nanoribbons described so far (Tr700). The spectra look similar but are different in detail. Most remarkable is the strong up shift of the C$_1$ line from 1245 to 1345 cm$^{-1}$, the downshift of the RBLM from 465 to 412 cm$^{-1}$, and the change of the line profile at the position of the G$^+$ line for the recording for the new object. Line positions for the new object are listed in Tab. 1, Col B$_n$.

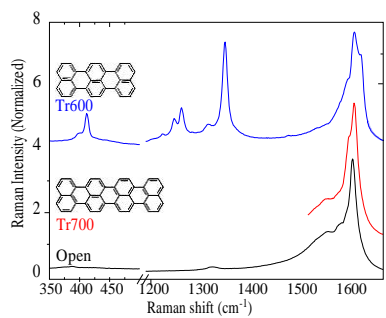

**Figure 5** Raman response for CNTs after transformation of the encapsulated ferrocene at temperatures of 700 (Tr700) and 600 °C (Tr600). The former spectrum was recorded for a 633 nm laser and the latter for a 568 nm laser. Both spectra are compared to the response after tube opening (Open). The left part of the abscissa depicts the R-BLMs. The small feature at 1245 cm$^{-1}$ in spectrum TR600 is from the nanoribbon already known. The inserts depict the chemical structure of terrylene and quaterrylene.

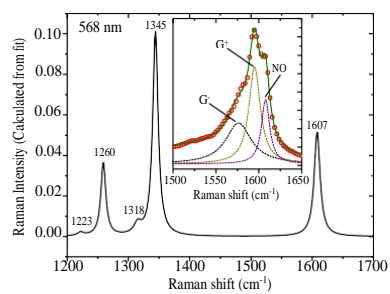

**Figure 6** Raman spectrum for the new object recorded with 568 nm laser after transformation at 600 °C. The spectrum was constructed from the values for the Raman lines from a fit to all lines in the spectrum. The insert depicts in particular the disentangling in the region around the G$^+$ line. The dots correspond to the experimental data. The strong line is the cumulative fit.

Due to the strong response of the NO its Raman spectrum could be extracted from a line fit to the response Tr600 in Fig. 5. This holds in particular also for the disentangling of the response around G$^+$ line of the tubes. Figure 6 depicts the spectrum as it was reconstructed from the values for the frequencies, line widths, and intensities of the fit. Due to the strong similarity of the spectra for the new object to the response of the nanoribbons we assign the object to nanoribbon 2 (NR2) in contrast to nanoribbon 1 (NR1) characterized by the C$_n$ lines.

**4 Discussion** The strong Raman response of the C$_1$ line is remarkable considering the fact that in its first observation in 2003 by W. Plank its intensity was less than 0.05 of the intensity of the G$^+$ line [11]. The combination of optimum tube diameter, transformation temperature, and excitation wave length is the reason for its dramatic increase. Now this line could even be observed as a very weak feature in off-resonance condition.

Transmission electron microscopy turned out to be difficult for the narrow tubes but were more clear for larger tubes as the latter are easier to disentangle. Since larger tubes tend also to develop inner tubes from the encapsulated ferrocene it is not surprising to see occasionally the ribbon structure in double-walled tubes. This is in agreement with the very weak observation of the Raman lines from the ribbons in such tubes. The distance of the white dots in Fig. 2 are consistent with an armchair configuration as it is expected for a quaterrylene-like molecule. The dependence of the outer tube and inner tube diameters on the orientation of the encapsulated transformation product is another strong evidence for its two-dimensional structure and thus supports our previously given interpretation as a quaterrylene-like feature or, more general, as fractions of graphene nanoribbons. The existence of nanoribbons inside SWCNT were reported previously several times [12, 13] although in these reports larger diameter tubes were used. Also, the growth of nanoribbons from polyaromatic hydrocarbons is well documented [8]. The dependence of the recorded tube diameter on time is evidence for the possibility of a diffuse rotation of the molecule which on the other hand is an indication that the lengths of the observed ribbons are rather short.

The observation of a precursor molecule after filling and mild transformation is interesting and surprising. Interesting, because it might be the reason why ferrocene was never observed in the Raman spectra after filling. Surprising, since temperatures of the order of 400 to 450 °C are really low considering the fact that 5-membered rings have to transform into 6-membered rings. Iron, which is still present in the tubes, may act as a catalyst. Also, very recently a transformation of pentagons into hexagon ribbons inside the tubes was shown from a computer simulation to be possible even at moderate high temperatures [14].

The observation of various temperature regions for the growth process indicates that different reactions dominate the growth process of NR1 in the different temperature regions. Nucleation is inhibited up to about 450 °C, though there is a small precursor molecule observable. This molecule has Raman pattern almost identical to quaterry-lene. It transforms quickly to the structure of NR1 which is strongly related to quaterrylene as we reported previously [5]. Region II is a nucleation region where more and more transformation products reach the full structure of the final product. For the temperature range III also tubes start to contribute where growth of nanoribbons is more difficult. Finally, at the very high temperatures the ribbons are not stable any more. They may transform into inner tubes which supports the observation that inner-tube growth does not decrease for the high transformation temperatures. Transformation of short pieces of nanoribbons (oligorylenes) inside SWCNT to an inner tube was recently observed explicitly [12].

From Tab. 1 the Raman lines of NR2 agree very well with the Raman lines of crystalline terrylene. This holds especially for the frequencies. The agreement is even better than the agreement of the lines for NR1 with the lines of polycrystalline quaterrylene. A noticeable discrepancy exists, however, for the RBLMs. In this case for both, the $C_n$ and the $B_n$ structures, the frequencies are considerably lower as compared to the frequencies in the crystalline forms. This may have to do with the fact that the breathing like modes certainly feel the confinement by the tubes much stronger and even serious changes of normal coordinates may occur.

With respect to intensities, agreement is less good. This is understandable, since resonance conditions depend strongly on normal coordinates of the vibrations which can be different for molecules outside and inside the tubes. Agreement between observed line positions and calculated frequencies is also very good, in particular for the fingerprint region. Less good agreement is obtained for the region of the C=C stretch modes. Here for both molecules the frequencies for the encapsulated molecules (1594, 1607 cm$^{-1}$) are noticeably higher than those calculated for the free molecules (1534, 1543 cm$^{-1}$), at least if one relates in all cases the lines assigned as "very strong" to the C=C stretch mode. On the other hand calculated transition energies and observed resonance positions agree very well for both nanoribbons. So, besides the discrepancy concerning the C=C stretch mode, terrylene-like molecules are a good suggestion for the structure of NR2.

The observation of a second species of NR suggests that there are several different types of NR growing inside the tubes. Whether they can be observed is a question of concentration and of finding the right laser energy to meet resonance conditions.

**5 Conclusion** In conclusion we demonstrate that Raman scattering is an excellent tool to reveal details of products grown inside SWCNTs. New objects are observed after filling with ferrocene and transformation at moderate high temperatures. The objects are found to be flat molecules from TEM and were characterized by Raman spectroscopy. They have all fingerprints of nanoribbons like quaterrylene and terrylene. For the new ribbon structure observed here for the first time a full Raman spectrum was extracted for the fingerprint region and including the C=C stretch modes and the radial breathing like modes.

**Acknowledgements** Work supported by the FWF project P21333-N20, by the OTKA project K115608, and by the EU project (2D-Ink FA726006). L.S. acknowledges the scholarship supported by the China Scholarship Council. Valuable discussions with P. Ruffieux are gratefully acknowledged.